\documentclass[%
 reprint,
superscriptaddress,
 amsmath,amssymb,
 aps,
pra,
floatfix,
]{revtex4-2}

\usepackage{graphicx}
\usepackage{dcolumn}
\usepackage{bm}
\usepackage{xcolor}
\usepackage{braket}
\usepackage{hyperref}
\usepackage[mathlines]{lineno}

\begin{document}

\preprint{APS/123-QED}

\title{Finite-Temperature Quantum Matter with Rydberg or Molecule Synthetic Dimensions}

\author{Sohail Dasgupta}
\email{sohail.dasgupta@rice.edu}
 \affiliation{Department of Physics and Astronomy, Rice University, Houston, TX 77005, USA}%
 \affiliation{Rice Center for Quantum Materials, Rice University, Houston, TX 77005, USA}%
\author{Chunhan Feng}%
\email{cfeng@flatironinstitute.org}
\affiliation{Center for Computational Quantum Physics, Flatiron Institute, 162 5th Avenue, New York, NY 10010, USA}

\author{Bryce Gadway}
\email{bgadway@illinois.edu}
\affiliation{Department of Physics, University of Illinois at Urbana-Champaign, Urbana, IL 61801-3080, USA}

\author{Richard T. Scalettar}
\email{scalettar@physics.ucdavis.edu}
\affiliation{Department of Physics, University of California, Davis, CA 95616, USA}%

\author{Kaden R. A. Hazzard}
\email{kaden.hazzard@rice.edu}
 \affiliation{Department of Physics and Astronomy, Rice University, Houston, TX 77005, USA}%
 \affiliation{Rice Center for Quantum Materials, Rice University, Houston, TX 77005, USA}%
 \affiliation{Department of Physics, University of California, Davis, CA 95616, USA}%

\date{\today}

\begin{abstract}
Synthetic dimension platforms offer unique pathways for engineering quantum matter. We compute the phase diagram of a many-body system of ultracold atoms (or polar molecules) with a set of Rydberg states (or rotational states) as a synthetic dimension, where the particles are arranged in real space in optical microtrap arrays and interact via dipole-dipole exchange interaction. Using mean-field theory, we find three ordered phases - two are localized in the synthetic dimension, predicted as zero-temperature ground states in Refs.~[Sci. Rep., 8, 1 (2018) and  Phys. Rev. A 99, 013624 (2019)], and a delocalized phase. We characterize them by identifying the spontaneously broken discrete symmetries of the Hamiltonian. We also compute the phase diagram as a function of temperature and interaction strength, for both signs of the interaction. For system sizes with more than six synthetic sites and attractive interactions, we find that the thermal phase transitions can be first or second order, which leads to a tri-critical point on the phase boundary. By examining the dependence of the tri-critical point and other special points of the phase boundary on the synthetic dimension size, we shed light on the physics for thermodynamically large synthetic dimension.  
\end{abstract}

\maketitle


\section{Introduction}\label{sec:intro}

Synthetic dimension platforms are more than a powerful tool for investigating interesting physics from broad fields, they are also a pathway for simulating interacting quantum matter that has no analog in other systems. A synthetic dimension is built using the internal or motional states of quantum particles such as ultracold atoms, molecules, or photons. When the levels are coupled with electromagnetic radiation, these platforms can be used to engineer Hamiltonians that are identical to a wide variety describing motion in real space. They are highly tunable, allowing independent control of the system parameters, including tunneling amplitudes and phases, and the on-site synthetic potentials. 

Since the first proposal \cite{boada2012quantum}, several platforms have been experimentally realized. Examples include synthetic dimensions based on nuclear spin states \cite{stuhl2015visualizing,  mancini2015observation, li2022bose}, momentum states \cite{gadway2015atom,an2017direct,an2017diffusive,meier2016atom,xiao2020periodic,an2021nonlinear,meier2016observation,meier2018observation,liang2022dynamic}, optical clock states \cite{livi2016synthetic}, harmonic trap states \cite{price2017synthetic, oliver2023bloch}, Floquet states \cite{Xu2022floquet} and Rydberg states \cite{signoles2014confined,kanungo2022realizing,chen2023strongly} of ultracold atoms; and time and frequency states of photons \cite{luo2015quantum, ozawa2016synthetic,ozawa2017synthetic,yu2023moire,lin2018}. There are also proposals to build synthetic dimensions using rotational states of polar molecules \cite{sundar2018synthetic}.  Observation of topological edge states \cite{ozawa2019topological,wang2015topological,meier2016observation,chen2018experimental,salerno2019quantized,cai2019experimental,an2017direct,zhang2018topological,mancini2015observation,kanungo2022realizing,chalopin2020probing,dutt2020higher}, Anderson localization \cite{an2017diffusive,meier2018observation}, non-linear physics \cite{an2021nonlinear} and the non-Hermitian  skin effect \cite{liang2022dynamic}, realization of synthetic gauge fields~\cite{celi2014synthetic,stuhl2015visualizing,  mancini2015observation}, and hyperbolic lattices~\cite{zhang2021efimov,yu2020topological} are some of the highlights. Proposals to engineer topological quantum field theory models~\cite{shen2022simulating} and other topological physics~\cite{yan2019emergent} further the versatility of these platforms.

Synthetic dimension platforms have been utilized to observe not only single-particle phenomena but also unique interacting physics. Alkaline-earth atoms with nuclear spin states interact via SU($N$) symmetric interactions that are non-local in the synthetic space \cite{mancini2015observation,barbarino2015magnetic,yan2015topological,barbarino2016synthetic,zeng2015charge}, photonic states interact via synthetic site preserving interactions \cite{ozawa2017synthetic}, momentum-space lattices have local (on-site) attractive interactions \cite{an2018correlated,an2018engineering}, and Rydberg atoms interact via dipolar exchange interactions \cite{chen2023strongly} that are local (roughly nearest-neighbor) in synthetic space. 

In this work, we focus on a model of dipolar interacting quantum many-body systems, first proposed to be built with rotational states of ultracold polar molecules trapped in optical microtraps \cite{sundar2018synthetic}. The same model can also be realized with Rydberg states of ultracold atoms \cite{ozawa2019topological, kanungo2022realizing, chen2023strongly}. 

One of the intriguing features of this system is that for strong interactions compared to the synthetic tunneling rates, the ground states are localized to finitely many sites in the synthetic dimension. They resemble thick strings (membranes) in one (two) dimensional real-space arrays of molecules or Rydberg atoms fluctuating in a two (three) dimensional real+synthetic space, hence named the string/membrane phase. Essential features of these phases have been studied for some special cases. In Ref. \cite{sundar2019strings}, the wavefunction for the string phase was exactly solved for one real and one synthetic dimension when the synthetic tunneling rates vanish. Mean-field theory \cite{sundar2018synthetic} and density matrix renormalization group (DMRG) \cite{sundar2019strings} predict that the string/membrane phase persists to infinitely large synthetic dimensions for any synthetic tunnelings for attractive interactions, while for repulsive interactions there is a critical synthetic tunneling at which the system transitions from a string/membrane to a  disordered phase. Recently, stochastic Green's function quantum Monte Carlo studies \cite{feng2022quantum} showed that in two-dimensional real-space arrays, the membranes survive to finite-temperatures in systems with a finite number of synthetic lattice sites, and undergo a thermal phase transition into the disordered phase as temperature is raised. 

However, fundamental questions remain unanswered: (1)~Are these the only phases of this model?  (2)~Is there a simple physical understanding of the phases? (3)~Do they persist at finite temperature for repulsive interactions?  (4)~How does the size of the synthetic dimension affect the phases and the critical temperature?  

In this paper, we make progress on these questions. We identify three symmetry-broken phases, accessible to experiments by tuning a single system parameter. Two of the phases are localized in the synthetic dimension and were predicted in Ref. \cite{sundar2018synthetic} and studied in \cite{sundar2019strings} and \cite{feng2022quantum}, while our calculations reveal an additional ordered delocalized phase.  Through identifying the symmetry groups and a mean-field theory, we classify them by the discrete symmetries that they break and show that all of the non-trivial phases persist to finite temperature. We characterize the dependence on the synthetic dimension size of all the phases and the transitions between them. 

We also find a tri-critical point on the thermal phase boundary for attractive interactions when there are more than six synthetic sites. In this case, the thermal transition is first order for weak synthetic tunnelings and second order for strong synthetic tunnelings with a tri-critical point in between. We discuss the analogy of this phenomenon with the classical Potts and $p$-states clock model later.

The paper is organized as follows: Section~\ref{sec:synth_dim_set} discusses the platform and the Hamiltonian, Sections~\ref{sec:mft} and \ref{sec:results} describe the mean field theory and the resulting phase diagram, and we conclude in Section~\ref{sec:conclustion} and suggest interesting open questions.

\section{Rydberg Atom and Molecule Synthetic Dimensions}\label{sec:synth_dim_set}
\begin{figure}[t]
    \centering
    \includegraphics[width=\linewidth]{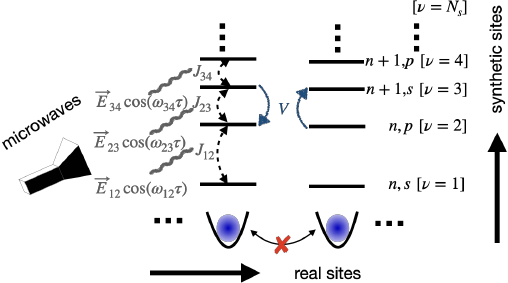}
    \caption{Two real-space sites of a Rydberg atom ($n>> 1$) array with $N_s$ synthetic sites. The synthetic sites are indexed by $\nu$ ($1$ to $N_s$). Resonant microwaves couple alternating $s$ (odd $\nu$) and $p$ (even $\nu$) angular momentum states. Atoms interact via dipolar exchange. Deep traps prevent real space tunneling.}
    \label{fig:int_synth_dim}
\end{figure}

We consider a system of ultracold Rydberg atoms or polar molecules, trapped individually in optical microtraps or sites of a optical lattice in a one- or two-dimensional bipartite real-space geometry (Fig. \ref{fig:int_synth_dim}). (Results for attractive interactions apply also to non-bipartite lattices.) Rydberg levels of atoms or rotational states of molecule are separated by microwave frequencies, allowing coherent control of several levels. The lattice is sufficiently deep to prevent real space tunneling. Throughout the rest of the paper, we describe the system in terms of Rydberg states of ultracold atoms but equivalent physics exists in polar molecules as well.

A set of Rydberg states, alternating between $s$ and $p$ angular momentum levels, is resonantly coupled with microwaves to form a linear synthetic dimension, as depicted in Fig. \ref{fig:int_synth_dim}. This forms an open boundary condition, or the last state can be coupled to the first to have a periodic boundary condition. The power of the microwaves coupling pairs of synthetic sites sets the corresponding tunneling amplitude.  The detuning of the microwaves from the resonant frequency sets the on-site potential energy. In our model, we fix all the synthetic tunnelings to be equal to each other and all the detunings to be zero, which is easy to achieve experimentally. 

Pairs of Rydberg atoms interact strongly via the dipole-dipole interaction \cite{Gallagher1994}. In general, interactions depend on the synthetic sites of the two atoms. But our choice of alternating $s$ and $p$ angular momentum states implies that the interaction can be non-zero only between sites of different parity with the pairs of states next to each other being the strongest. For most scenarios, the interaction strength between sites separated by three synthetic sites are small and is ignored in our calculations. The interaction strength does not vary significantly with the principal quantum numbers $n$ of the Rydberg states for large $n$. Hence we approximate to having non-zero interaction strengths only when two atoms are separated by exactly one synthetic site and assume them to be equal.

The Hamiltonian describing the situation is
\begin{equation}\label{eq:ham_res}
\begin{aligned}
H = &-J\sum_{\nu=1}^{N_s}\sum_i \ket{\nu,i}\bra{\nu-1,i} + h.c. \\ &+V\sum_{\nu=1}^{N_s}\sum_{\braket{i,j}}  \ket{\nu,i;\nu-1,j}\bra{\nu-1,i;\nu,j} + h.c.,
\end{aligned}
\end{equation}
where $N_s$ is the total number of synthetic sites, $J$ is the tunneling amplitude between a pair of synthetic sites, which we assume throughout to be positive, $V$ is the interaction energy between two atoms, and $\ket{\nu,i}$ represents an atom at real (synthetic) site $i$ $(\nu)$ with $\nu=0$ and $\nu=N_s$ identified to induce periodic boundary condition. We have assumed the quantization axis is perpendicular to the array (or otherwise oriented so that the dipolar interaction is isotropic), and have truncated the $1/r^3$ interaction to nearest neighbors. In 1D we expect this truncation to have only relatively small effects on the phase diagram, and this is likely also true in 2D. In 3D the $1/r^3$ interaction and anisotropy are likely to play major roles, a question we leave to future work. Some further discussion on these points can be found in the conclusions.

Considering an infinitely large lattice in real space and periodic boundary condition in the synthetic dimension, the Hamiltonian of Eq. \ref{eq:ham_res} has a $\mathcal{D}_{2N_s}\times \mathcal{T}$ symmetry; $\mathcal{D}_{2N_s}$ is the symmetry group of a regular polygon of $N_s$ sides and $\mathcal{T}$ is the discrete symmetry group of the real space lattice. 

\section{Mean Field Theory}\label{sec:mft}
To calculate the phase diagram of the Hamiltonian in Eq. \ref{eq:ham_res}, we employ a mean-field approximation to decouple the interaction terms, giving
\begin{equation}\label{eq:mean_field_ham_red}
\begin{aligned}
\mathcal{H}= &\sum_{\nu=1}^{N_s} \sum_{s=\pm 1} \left(-\frac{J}{2}+\tilde{V}\phi^*_{\nu,\bar{s}}\right)\ket{\nu,s}\bra{\nu-1,s} + h.c. \\ 
\end{aligned}
\end{equation}
where $s=\pm 1$ label the two sublattices of the bipartite lattice, $\bar{s}=-s$, $\tilde{V}=\frac{Vz}{2}$ with $z$ being the number of nearest neighbor atoms, and 
\begin{equation}\label{eq:mean_fields}
    \phi_{\nu,s} = \Big\langle\ket{\nu,s}\bra{\nu-1,s}\Big\rangle
\end{equation}
are the mean-fields. As mentioned above, periodic boundary condition is imposed by identifying $\nu=0$ and $\nu=N_s$. We allow the mean-fields to differ on the two different sublattices in the real dimension. This will be necessary to describe the phases for $V>0$. Expanding the expectation value on the right hand side of Eq. \ref{eq:mean_fields}, we see that the mean-fields satisfy the self-consistent equation
\begin{equation}\label{eq:mft_self_consistent}
    \phi_{\nu,s} = \frac{1}{\mathcal{Z}}\sum_\alpha \braket{\alpha\ket{\nu,s}\bra{\nu-1,s}\alpha}e^{-E_{\alpha}/T},
\end{equation} 
where $\mathcal{Z} = \sum_\alpha e^{-E_\alpha/T}$ is the mean-field partition function, and $\ket{\alpha}$ and $E_\alpha$ are the $\alpha^{\text{th}}$ eigenstate and eigenenergy of the mean-field Hamiltonian in Eq.~\ref{eq:mean_field_ham_red}, respectively.
Note that $\ket{\alpha}$ and $E_\alpha$ are functions of the mean-fields, $\{\phi_{\nu,s}\}_{\nu=1,\cdots,N_s;s=\pm 1}$. We have set the Boltzmann constant, $k_B=1$. 

The mean-field approximation identifies all real lattice points in a sublattice, thus reducing the symmetry group $\mathcal{T}$ to $\mathbb{Z}_2$. Therefore, the mean-field Hamiltonian (Eq. \ref{eq:mean_field_ham_red}) has $\mathcal{D}_{2N_s}\times \mathbb{Z}_2$ symmetry when all the mean-fields are equal, which corresponds to the disordered phase.  

We iteratively solve for the mean-fields using Eq. \ref{eq:mft_self_consistent}. We start with a random initial seed for each mean field in $(-1,1)$. The results are considered converged when $|\phi_{\nu,s}^k - \phi_{\nu,s}^{k-1}|<10^{-6}\ \forall \nu,s$; $k$ labels the iteration step.  We estimate a relative error of less than $2\%$, where the relative error is defined as  $|(\phi_{\nu,s}^{k_{\text{max}}}-\phi_{\nu,s}^{k_{\text{max}}/2})/\phi_{\nu,s}^{k_{\text{max}}}|$, where $k_{\text{max}}$ is the number of iterations needed for the mean-fields to be considered converged.

\section{Phase Diagram}\label{sec:results}
\begin{figure*}[ht]
    \centering
    \includegraphics[width=\linewidth]{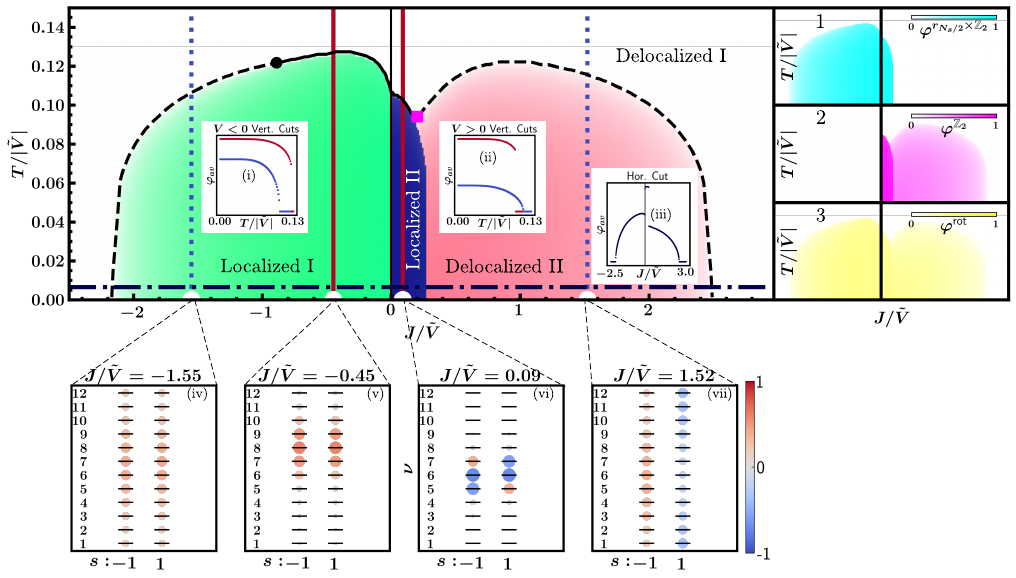}
    \caption{ Phase diagram for $N_s=12$ in the $T/\tilde{|V|}-J/\tilde{V}$ plane.  The green, blue, and red regions correspond to ``Localized I",  ``Localized II" and   ``Delocalized II" phases respectively and the white area corresponds to the disordered phase, ``Delocalized I". On the right, panels 1, 2, and 3 show the density plots of the three order parameters, $\varphi^{r_{N_s/2}\times \mathbb{Z}_2}$, $\varphi^{\mathbb{Z}_2}$, and  $\varphi^{\text{rot}}$. The x and y axes in the panels 1, 2, and 3 are identical to the main figure. The phase diagram is obtained by superposing the three order parameters, colored as in panels 1, 2, and 3, so that each phase is uniquely identified. The black solid (dashed) line demarcates the phase boundaries corresponding to first (second) order transitions between the ordered and the disordered phases. The black circle (magenta square) on the phase boundary denotes the tri-critical point (meeting point of Localized II, Delocalized II, and Localized I phases). Insets (i) and (ii): The average of the three order parameters, $\varphi_{\text{av}}=\frac{1}{3}(\varphi^{r_{N_s/2}\times \mathbb{Z}_2} + \varphi^{\mathbb{Z}_2} + \varphi^{\text{rot}})$ versus $T/|\tilde{V}|$; blue squares (red dots) correspond to the blue dotted (red solid) vertical cut.  (iii) The average of the order parameters is plotted versus $J/\tilde{V}$ for the dark blue dash-dot horizontal cut.  Callouts (iv)-(vii): The ground state wavefunction of the mean-field Hamiltonian at $T=0$ corresponding to $J/\tilde{V}$ at the white semi-circles. The y-axis corresponds to the $N_s$ discrete synthetic sites, labelled by $n$. The columns correspond to the two sublattices, $s=\pm 1$. Both the color intensity and marker size vary proportionally with the absolute value of the wavefunction.}
    \label{fig:phase_diagram_full}
\end{figure*}

\begin{figure}[t]
    \centering
    \includegraphics[width=\linewidth]{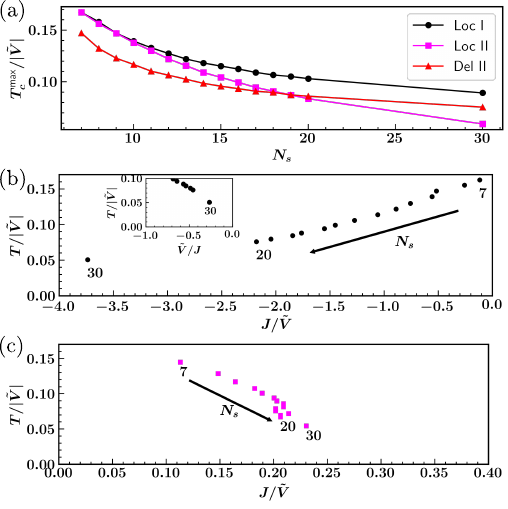}
    \caption{Trends of the phase diagram with $N_s$. (a) Scaling of the highest critical temperature in each of the three phases with $N_s$. It appears that all three of the $T_c^{\text{max}}$ scale to $0$ with $N_s$ although at different rates.(b) Scaling of the tri-critical point with $N_s$. The plot suggests that the tri-critical point scales to $J^{N_s=\infty},T^{N_s=\infty} = -\infty,0$. This is even more prominent in the inset, where the x-axis is inverted to $\tilde{V}/J$. (c) Scaling of the meeting of the Localized II, Delocalized I and Delocalized II phases with $N_s$. The plot suggests that  it scales to zero temperature but a finite tunneling as $N_s\to\infty$.  }
    \label{fig:scaling}
\end{figure}

Mean-field theory predicts a phase diagram with four phases (Fig.~\ref{fig:phase_diagram_full}). They are characterized by the subgroups of $\mathcal{D}_{2N_s}\times \mathbb{Z}_2$, the symmetries of the Hamiltonian in Eq.~\ref{eq:mean_field_ham_red}. The Localized I phase (green) breaks the ${\mathcal D}_{2N_s}$ symmetry but retains the ${\mathbb Z}_2$ symmetry, while the Localized II  (blue) phase breaks both.  The Delocalized II (red) phase preserves only the symmetry of simultaneous rotation by half the synthetic dimension size and real-space exchange. The Delocalized I (white) phase is the trivial or disordered phase where all the mean-fields are equal and the system has the full $\mathcal{D}_{2N_s}\times \mathbb{Z}_2$ symmetry. Callouts (iv)-(vii) of Fig.~\ref{fig:phase_diagram_full} depict the ground state wavefunctions in different phases, corresponding to the white semi-circles in the main figure. 

The phases are diagnosed with three order parameters,
\begin{align}
    \varphi^{\text{rot}}_{s} &= \left|\sum_{\nu=1}^{N_s} \phi_{\nu,s} e^{-\iota 2\pi \nu/N_s} \right | \label{eq:ord_par_rot}\\
    \varphi^{\mathbb{Z}_2} &= \sum_{\nu=1}^{N_s} \left |\phi_{\nu,1} -\phi_{\nu,-1} \right |^2 \label{eq:ord_par_z2} \\
    \varphi^{r_{N_s/2}\times\mathbb{Z}_2} &= \sum_{\nu=1}^{N_s} \left |\phi_{(\nu+N_s/2),1}-\phi_{\nu,-1} \right |^2 \label{eq:ord_par_rn2z2}
\end{align}
for even $N_s$. Each vanishes if and only if the corresponding symmetry is preserved. $\varphi^{\text{rot}}_s$ distinguishes phases with from those without all the rotation symmetries of the dihedral group, $\mathcal{D}_{2N_s}$. The value of $s$ is irrelevant for the phases in the mean-field theory. From Fig. \ref{fig:phase_diagram_full}.3, it is evident $\varphi^{\text{rot}}$ separates the trivial phase, Delocalized I, from the rest of the phases. $\varphi^{\mathbb{Z}_2}$ distinguishes phases with sublattice exchange symmetry. The sublattice exchange symmetry, characterized by $\varphi^{\mathbb{Z}_2}$, is never broken for attractive interactions as seen in Fig. \ref{fig:phase_diagram_full}.2 but is broken in all the non-trivial phases for repulsive interactions.  For even $N_s$, the Delocalized II phase (red region Fig. \ref{fig:phase_diagram_full}) is the only non-trivial phase preserving the $r_{N_s/2}\times \mathbb{Z}_2$ symmetry. For odd $N_s$, rotation by $N_s/2$ is not well-defined. Instead if we perform rotation by $(N_s\pm 1)/2$ sites, then $r_{(N_s\pm 1)/2}\times\mathbb{Z}_2$ is ``almost" a symmetry. The order parameter in Delocalized II phase is small ($<0.1$) compared to Localized I/II phases ($0.3-1$) but not exactly zero. The phase diagram of Fig. \ref{fig:phase_diagram_full} is sketched by adding the three different colors (in an RGB sense) of Fig. \ref{fig:phase_diagram_full}.1 - \ref{fig:phase_diagram_full}.3 at individual points.

In addition to their symmetry-breaking features, an intriguing feature of the Localized I and II phases is that they are localized along finitely many synthetic sites (i.~e., the probability decays exponentially outside of a finite number of synthetic sites) as seen for the ground states shown in the callouts (v) and (vi) of Fig.~\ref{fig:phase_diagram_full}. The existence of two localized ground states was already predicted in Ref.~\cite{sundar2018synthetic} using mean-field theory. For $J=0$, Ref.~\cite{sundar2019strings} analytically proved that one-dimensional real-space arrays have ground states confined to two or three adjacent synthetic sites.   Similar behavior is  observed in our $J=0$ mean-field theory: width-2 and width-3 states are degenerate, and lower in energy than the disordered phase when $N_s>4$, as shown in Appendix \ref{app:analytics}. 
 
The Delocalized II phase, not reported in Refs.~\cite{sundar2018synthetic,sundar2019strings}, is spread over a large fraction of the synthetic dimension, but the atoms are spread over opposite sets of synthetic sites on each sublattice. The system however remains invariant under simultaneous rotation by half of the synthetic sites and real space sublattice exchange for even number of synthetic sites.

All three non-trivial phases extend to non-zero temperatures for finite $N_s\geq 5$. However, the transition temperatures tend to zero as $N_s\to \infty$. A Peierls-like argument comparing the free energies, $\Delta\mathcal{F}=\mathcal{F}_{\text{ordered}}-\mathcal{F}_{\text{Delocalized I}}$, of the ordered phases and the Delocalized I phase suggests this, and it is consistent with our analysis of the critical temperatures' dependence on $N_s$. We plot the highest critical temperature in each of the three ordered phases as a function of $N_s$ in Fig.~\ref{fig:scaling}(a).

The orders of the thermal phase transitions depend on $N_s$ and the ratio $J/\tilde{V}$. 
The phase boundary between the Localized I and Delocalized I phases hosts a tri-critical point for $N_s>6$. We observe that the phase transition changes from being first-order at small $J/|\tilde{V}|$ to being second-order at large $J/|\tilde{V}|$ with a tri-critical point in between (black circle in Fig.~\ref{fig:phase_diagram_full}). The value of $T/|\tilde{V}|$ ($J/|\tilde{V}|$) at the tri-critical point decreases (increases) with $N_s$ [Fig.~\ref{fig:scaling}(b)]. The inset in Fig.~\ref{fig:scaling}(b) strongly suggests that in the limit $N_s\to\infty$, the tri-critical point is at $T/|\tilde{V}|=0,J/\tilde{V} = -\infty$. This is consistent with the arguments of Ref.~\cite{sundar2018synthetic} that the $V<0$ ground state for $N_s\to\infty$ is always a string.

We  understand the qualitative behavior of the tri-critical point in two ways: an analytic argument for $J=0$ and analogies to previously studied classical models. In App.~\ref{app:analytics}, we analytically show that the thermal phase transition is first (second) order at $J=0$ for $N_s > 6$ ($N_s\le 6$).  Furthermore, we note that the $N_s$-dependence of the order of the phase transition resembles that of the classical Potts model \cite{ortiz2012dualities}, if the synthetic sites are considered as spin indices of a large spin $[S=(N_s-1)/2]$ system. For example, for the two-dimensional Potts model, the thermal phase transition is second (first) order when $N_s \le 4$ ($N_s >4$). For another large-spin model, the clock model, the analogy is less direct: the system undergoes a second order thermal phase transition when $N_s \le 4$, and has a Berezenskii-Kosterlitz-Thouless (BKT) phase intermediate to ordered and disordered phases for $N_s > 4$. 

In making the analogy to classical models, it's worth noting that the synthetic dimension Hamiltonian  Eq.~\ref{eq:ham_res} has qualitative differences with both Potts and clock models. The Potts model has a different symmetry -- a full spin-permutation symmetry, whereas the clock model has the same $D_{2N_s}$ symmetry but has nonlocal interactions between all synthetic sites. 

A model that captures both the symmetry and the locality of the interactions in synthetic space was introduced and studied in Ref.~\cite{cohen2022classical}. It includes same-synthetic site ($J_0$) and nearest-synthetic-site ($J_1$) interactions. When $J_1 \gg J_0$, the naïve case corresponding to the quantum model, a direct phase transition from a narrow sheet (dubbed ``ferromagnetic" there) to the disordered phase is found. For large $N_s$ this is clearly first order, while for smaller $N_s$, the order is less clear -- it is either a less drastic first order or second order transition (e.g. Fig.~3 of Ref.~\cite{cohen2022classical}).  This classical analogy also suggests other interesting phenomena. For example, when $J_1$ is sufficiently bigger than $J_0$ the transition becomes second order and may involve a second crossover or Berezinskii-Kosterlitz-Thouless transition, an interesting possibility in the quantum model.

The $N_s$-dependence of the meeting point of the Localized II, Delocalized II, and Delocalized I phases for repulsive interactions shows that the Localized II phase extends only to finite $J/\tilde{V}$ even as $N_s\to\infty$, in contrast with the Localized I phase [Fig.~\ref{fig:scaling}(c)]. From Fig.~\ref{fig:scaling}(c), it seems that the $T/\tilde{V}$ goes to zero, but $J/\tilde{V}$ tends to a finite value as $N_s\to\infty$. This qualitatively agrees with the findings of Ref.~\cite{sundar2018synthetic} that for $N_s\to\infty$ and repulsive interactions, the system undergoes a quantum phase transition. 
Our phase boundary for attractive interactions is also qualitatively consistent with that computed with stochastic Green's function (SGF) quantum Monte Carlo (QMC) \cite{feng2022quantum}. The SGF-QMC method is sign-problem free for attractive interactions (only) and thus can calculate observables to high precision. Fig.~9 of Ref.~\cite{feng2022quantum} shows the phase diagram for $N_s=10$ for a 2D square lattice, finding a finite-temperature phase transition between the quantum membrane (Localized I) and disordered (Delocalized I) phases, as we predict from mean-field theory.  The SGF-QMC transition temperature is similar to, though shifted from the mean-field predictions, and the trends with $N_s$ and $J/\tilde{V}$ are similar.  For example, both the QMC and the mean-field theory predict that the system undergoes a $T=0$ quantum phase transition at finite $J/\tilde{V}$; for $N_s=10$, QMC finds this transition to be at $J/\tilde{V}\approx-1.2$, whereas our mean-field results determine it at $J/\tilde{V}\approx -1.7$. Mean-field theory predicting the quantum critical point at a larger $J/\tilde{V}$ is unsurprising, since it ignores all fluctuations.  In addition, QMC observes only second-order phase transitions for $N_s=10$.  This could either be because the tri-critical point for $N_s=10$ is very weakly first order, because it occurs at a much lower value of $J/\tilde{V}$ than shown in the QMC studies, or it could be that the transitions at all $J/{\tilde V}$ for $N_s=10$ are second order. In the latter case, it is possible that the critical $N_s$ above which the tri-critical point first appears has been shifted from the $N_s=6$ value predicted by mean-field theory to $N_s>10$. We believe this is likely, as the QMC results for $N_s=14$ and $V=-5J$ (Fig.~7 of Ref.~\cite{feng2022quantum}) show a step feature suggestive of a first-order transition.

\section{Conclusion and Outlook}\label{sec:conclustion}
 We calculated the phase diagram of a dipolar interacting quantum system with real and synthetic dimensions, and analyzed the features of the phase diagram as a function of interaction, temperature, synthetic tunnelings, and synthetic dimension size. This model can be engineered with ultracold Rydberg atoms or polar molecules arranged in optical microtraps or optical lattices and external microwave couplings, with a recent first study of its dynamics in Ref.~\cite{chen2023strongly}. Tuning a single parameter, $J/\tilde{V}$, realizes a rich phase diagram with four distinct phases, and both thermal and quantum phase transitions, which may be either first or second order.

Using mean field theory and analyzing the results according to order parameters constructed based on the symmetry group, we classify the string/membrane phases, predicted in Refs.~\cite{sundar2018synthetic,sundar2019strings}.  Both these phases spontaneously break the $\mathcal{D}_{2N_s}$ symmetry of the model, with some remnant symmetries depending on the sign of $V$ and whether $N_s$ is even or odd. For $V>0$, this phase breaks the real-space sublattice-exchange ${\mathbb Z}_2$ symmetry.

In addition, we predict the existence of another symmetry-broken phase that is not string/membrane-like. This phase occurs for $V>0$ over an intermediate range of $J/\tilde{V}$, and it enjoys a remnant symmetry under a simultaneous rotation by half the synthetic sites and sublattice exchange (for even $N_s$). Moreover, the system is delocalized over a finite fraction of the synthetic space. It is an open question to what extent the localization plays a role beyond symmetry-breaking-- are the universal properties of the low-energy excitations of the ordered phases solely determined by the symmetry, or does the stringiness also play some additional role? 

We observe the presence of both first- and second-order transitions with a tri-critical point in between for $N_s>6$ and $V<0$. The thermal phase transitions are second order for $N_s\leq 6$.  The dependence of the order of the transition on $N_s$ resembles Potts model physics.

The scaling of the phase boundaries with $N_s$ is consistent with earlier $T=0$ mean-field calculations and the special cases that have been treated numerically. We show that the tri-critical point scales to $(J/\tilde{V}\to-\infty, T\to 0)$ as $N_s\to\infty$; hence the Localized I phase persists to arbitrarily large $|J/\tilde{V}|$ at $T=0$ for $N_s\to\infty$. Similarly, we show the meeting point of the Localized II, Delocalized II and Delocalized I phases scale to a finite $J/\tilde{V}$ and zero $T$ with $N_s\to\infty$. These results are qualitatively consistent with previous predictions \cite{sundar2018synthetic,sundar2019strings} that the string phase for $V>0$ $(V<0)$ persists to finite (arbitrarily large) $J/\tilde{V}$.

The physical realization of the different phases and observation of the phase transitions appears within the reach of current experiments, for example using the Rydberg tweezer platform of Ref.~\cite{chen2023strongly}. 
Usually, in experiments it is convenient to start with a product state, \textit{e.g.}, $\ket{\psi}=\ket{1111\cdots}$. The $T=0$ ground state for different $J/\tilde{V}$ values can be adiabatically prepared by slow variations of the microwave parameters (detunings and amplitudes).
Importantly, to capture features of the (De)Localized II phases one also needs to break the spatial translation and reflection symmetry during state preparation, as has been demonstrated in Ref.~\cite{Chen2023} for a special $N_s=2$ case.
Following adiabatic preparation, the properties of the ground state can then be explored to identify the different phases discussed. 
For example, measurements of the counting statistics-- measuring  the number of Rydberg atoms in a chosen level in a given shot, and then taking a histogram of this quantity--  can provide a means to distinguish between the Localized and the Delocalized phases. Delocalized phases would have broad, roughly Poissonnian,  fluctuations around a mean value, whereas localized phases would have a bimodal character: a peak with $O(1/N_s)$ probability of seeing around $O(N_s)$ Rydberg atoms in a given level, and an O(1) peak around having no particles in the level. Alternatively, by attempting to ramp from and then back to the state $\ket{1111\cdots}$, one can identify accessible phase transitions by finding breakdowns in adiabaticity at particular $J/\tilde{V}$ values.

Effects of long-ranged interactions and interactions that vary with synthetic site on the phases are left for future investigations. For the dipolar real-space interactions, states that preserve sublattice symmetry will remain mean-field eigenstates with the same eigenenergy, where their only consequence is to change the effective $z$ factor in ${\tilde V}$. For states breaking the sublattice symmetry, the intra-sublattice (``A-A") versus inter-sublattice (``A-B") interactions will couple to different mean fields. Although a weak dependence of the interactions on the synthetic site index itself should not drastically change the phase diagram (Localized-Delocalized phase transitions change to sharp cross-overs), having interactions non-local in the synthetic space could have non-trivial effects.  

In the future, we expect the dipolar synthetic dimension platform to be a powerful tool for studying the interplay of interactions, topology and synthetic gauge fields. Incorporating different synthetic tunneling schemes or geometries along the synthetic dimension is experimentally feasible, so the already rich physics observed here is likely to be the tip of an iceberg.  Different tunneling schemes can be implemented by simply adjusting the power of the microwaves similar to the schemes in Ref.~\cite{suszalski2016different}. The Su-Schreiffer-Heeger (SSH) model, which hosts topological edge states, has been already realized in a single Rydberg atom \cite{kanungo2022realizing} and systems with gauge fields have been realized in the  platform of Rydberg atom tweezer arrays \cite{chen2023strongly}.

\begin{acknowledgments}
The authors thank Charles Dyall, Tao Chen, Chenxi Huang, Jacob Covey, Barry Dunning, and Thomas Killian for the helpful conversations. This work was supported in part by the Welch Foundation (C-1872), the National Science Foundation (PHY1848304), and the W. M. Keck Foundation (Grant No. 995764), and K.~R.~A.~H. benefited from
discussions at the KITP, which is supported in part by the National Science Foundation (PHY1748958), and the Aspen Center for Physics, which is supported in part by the National Science Foundation (PHY-1066293). The Flatiron Institute is a division of the Simons Foundation. The work of R.~T.~S. was supported by the U.S. Department of Energy, Office of Science, Office of Basic Energy Sciences, under Award Number DE-SC0014671.
B.~G. acknowledges support by the National Science Foundation under grant No.~1945031.
This work was supported in part by the Big-Data Private-Cloud Research Cyberinfrastructure MRI-award funded by NSF under grant CNS-1338099 and by Rice University's Center for Research Computing (CRC).

K.~R.~A.~H., R.~T.~S. and B.~G. planned the research. S.~D. and K.~R.~A.~H. performed the mean-field theory calculations. All authors interpreted results, compared with prior results (SGF-QMC, classical models, and similar), and evaluated experimental feasibility. S.~D. primarily wrote the initial draft and all the authors helped in writing the manuscript.
    
\end{acknowledgments}

\bibliography{apssamp}

\appendix

\section{$J=0$ Analytic Solution}\label{app:analytics}
When $J=0$, we can derive a set of simplified equations for the mean-fields and infer that the thermal phase transition is first order for $N_s>6$ and second order otherwise. In this limit, the mean-field Hamiltonian is
\begin{equation}\label{eq:mft_J0}
    \mathcal{H}= \sum_{\nu} \sum_{s=\pm 1} \tilde{V}\phi^*_{\nu,\bar{s}}\ket{\nu,s}\bra{\nu-1,s} + h.c.
\end{equation}
Numerically solving the mean-field equations, we find two solutions, a disordered phase in which all mean-fields are equal, and a string/membrane phase where one or two adjacent mean-fields are non-zero. Using this information as an ansatz, an analytic understanding of the phases and phase transitions can be obtained. For $N_s>4$, the mean-field ground state is the string/membrane state. For $N_s=4$, they are equal in energy and our mean-field theory cannot distinguish the two. 

We can predict the critical temperature and the order of the phase transition of the system at $J=0$ by considering the self-consistency equation for the mean-fields,
\begin{equation}\label{eq:tilde_ord_param_self_consistent}
    \tilde{\phi}_s = \frac{1}{2}\left(\frac{e^{-\beta \tilde{V}\tilde{\phi}_{\bar{s}}}-e^{\beta \tilde{V}\tilde{\phi}_{\bar{s}}}}{N_s -2 +e^{-\beta \tilde{V}\tilde{\phi}_{\bar{s}}}+e^{\beta \tilde{V}\tilde{\phi}_{\bar{s}}} }\right),
\end{equation}
where $\tilde{\phi_s} = \sum_{\nu=1,2} |\phi_{\nu,s}|^2$, $s$ labels the sublattice, $\bar{s}=-s$, and $m$ sums over the two potentially non-zero mean fields, which we have labeled as $1$ and $2$.
For small $\tilde{\phi}_{s}$,
\begin{equation}\label{eq:tilde_taylor}
\begin{aligned}
    \tilde{\phi}_s = &\left(\frac{\tilde{V}}{N_sT}\right)^2 \tilde{\phi}_s \\
    &+ (N_s-6)\tilde{V}^4\frac{((N_sT)^2+\tilde{V}^2)}{6N_s^5T^6}\tilde{\phi}_s^3 + \mathcal{O}(\tilde{\phi}_s^5)
\end{aligned}
\end{equation}
A phase transition occurs when the coefficient of the linear term, $1-(\frac{{\tilde V}}{N_s T})^2$,  vanishes and the order of the phase transition is determined by the sign of the $\tilde{\phi}_s^3$ term. This is equivalent to the standard Landau theory analysis of phase transitions with a $U(1)$ symmetry associated with the phase of ${\tilde \phi}_s$, with the coefficients determined using the mean-field theory. Thus, the critical temperature is $T_c=\tilde{V}/N_s$ when the transition is second-order. In Eq.~\ref{eq:tilde_taylor}, the coefficient of the $\tilde{\phi}_s^3$ term on the rhs is positive for $N_s>6$ and thus has a first-order transition, and negative for $N_s< 6$ and thus has a second-order transition.

\end{document}